\documentclass[lettersize,journal,onecolumn]{IEEEtran}
\usepackage{amsmath,amsfonts}
\usepackage{algorithmic}
\usepackage{array}
\usepackage[caption=false,font=normalsize,labelfont=sf,textfont=sf]{subfig}
\usepackage{textcomp}
\usepackage{stfloats,cite}
\usepackage{url}
\usepackage{verbatim}
\usepackage{authblk}

\usepackage{tabularx}
\usepackage{multirow}
\usepackage{graphicx,xcolor}

\def\BibTeX{{\rm B\kern-.05em{\sc i\kern-.025em b}\kern-.08em
    T\kern-.1667em\lower.7ex\hbox{E}\kern-.125emX}}
\usepackage{balance}
\begin{document}
\title{Deep Learning-Based Snow Depth Retrieval Using Sentinel-1
Repeat-Pass InSAR}

\author[1]{Nayan Yadav}
\author[2]{Shadi Oveisgharan}
\author[1]{Shirin Jalali}

\affil[1]{Department of Electrical and Computer Engineering, Rutgers University, Piscataway, NJ, USA}
\affil[2]{Jet Propulsion Laboratory, California Institute of Technology, Pasadena, CA, USA}



\maketitle

\begin{abstract}
Snow depth plays a central role in seasonal snowpack characterization and the terrestrial water cycle, yet remains challenging to estimate at high spatial resolution. Recent studies have shown that repeat-pass interferometric synthetic aperture radar (InSAR) measurements combined with physics-based models can enable effective snow water equivalent (SWE) retrieval. However, the performance of these methods depends strongly on measurement accuracy and modeling assumptions.

Building on the success of InSAR-based approaches, we develop a robust learning-based model that directly learns the relationship between measured InSAR observables and snow depth. The model is trained on a single SnowEx Idaho site and evaluated across independent years and geographically distinct regions. Results demonstrate strong temporal and spatial transferability. In temporal transfer experiments, the proposed approach achieves a Pearson correlation of 0.81 with lidar snow depth, compared to a correlation of approximately 0.47 reported for physics-based Sentinel-1 SWE retrievals over the same site.
\end{abstract}


\section{Introduction}

Global water supply for billions of people depends strongly on seasonal snowpacks
\cite{barnett2005potential}. In snow-dominated watersheds, snow accumulation and melt regulate streamflow timing, groundwater recharge, and downstream water availability, making accurate characterization of snow storage essential for hydrologic forecasting and water-resource management
\cite{li2017much,lorenzi2024tracking}. This need is becoming more urgent as climate change alters snowfall fraction, snow persistence, and the elevation dependence of snow accumulation, reducing the predictability of snowmelt-driven runoff in many basins
\cite{klos2014extent,mccrystall2021new,siirila2021low}.

Despite its importance, high-resolution mapping of snow depth and snow water equivalent (SWE) remains difficult. Ground-based networks provide valuable time series but have limited spatial representativeness in complex terrain, while airborne lidar can measure snow depth accurately at fine spatial resolution but is costly, episodic, and sensitive to acquisition conditions
\cite{dozier2016estimating,painter2016airborne}. Satellite remote sensing  provides a critical pathway toward large-scale snow monitoring, especially in mountainous regions where snow exhibits strong spatial heterogeneity.

Recent advances in satellite radar missions have renewed interest in employing  synthetic aperture radar (SAR) and interferometric SAR (InSAR) for snow characterization. Empirical Sentinel-1 approaches have shown that radar observables can support snow-depth retrieval at sub-kilometer resolution over mountain ranges
\cite{lievens2019snow,lievens2022sentinel}. Physics-based interferometric methods have demonstrated that repeat-pass phase measurements are sensitive to changes in dry-snow SWE, providing a physically grounded route to snow-mass retrieval
\cite{guneriussen2002insar,leinss2015snow,eppler2022snow,oveisgharan2024snow}. These results highlight the promise of InSAR for high-resolution snow monitoring, but they also underscore a practical limitation: retrieval accuracy depends strongly on interferometric phase quality, coherence, phase unwrapping, and simplifying assumptions about snow and scattering conditions
\cite{zebker1992decorrelation,eppler2022snow}.

Machine learning offers a complementary alternative in which the relationship between measured observables and snow state is learned directly from data rather than imposed through an explicit inversion model. A recent study showed that, under the assumption that seasonal snow-depth patterns remain spatially consistent, learning-based models can combine InSAR observables with terrain and ancillary variables to estimate snow depth under site-specific conditions over relatively flat terrain
\cite{alabi2025advancing}. In this letter, we extend that idea to a more challenging setting and develop a compact learning-based model that directly maps measured InSAR observables to snow depth in mountainous terrain. Trained using data from a single SnowEx Idaho site, the model is evaluated across independent years and geographically distinct regions. The results show that the learned relationship transfers across both time and space, indicating that useful snow-depth information can be extracted from InSAR measurements without relying on site-specific spatial pattern stability or a fully specified physics-based retrieval framework.

\begin{figure*}[!t]
\centering
\subfloat[]{%
  \includegraphics[width=0.27\textwidth]{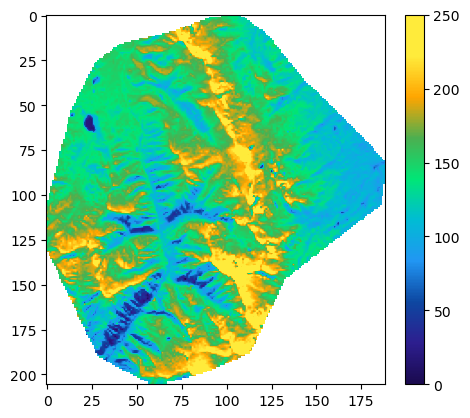}%
}
\subfloat[]{%
  \includegraphics[width=0.27\textwidth]{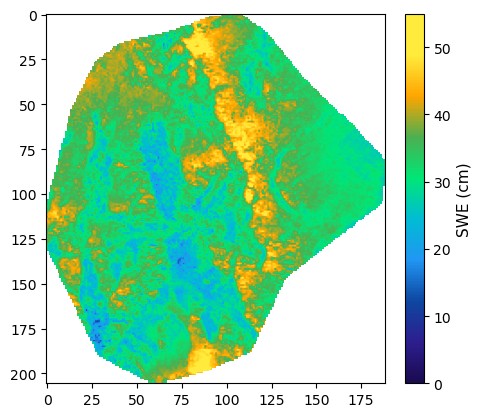}%
}
\subfloat[]{%
  \includegraphics[width=0.27\textwidth]{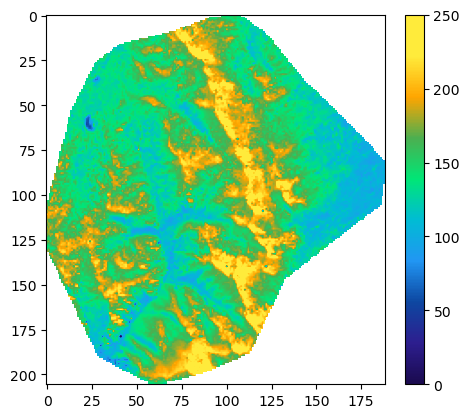}%
}

\caption{Temporal transfer results for BS2020 $\rightarrow$ BS2021. Panels show (a) lidar snow depth reference, (b) Retrieved SWE using physics-based approach \cite{oveisgharan2024snow}, and (c) MLP-predicted snow depth.}
\label{fig:bs2021_transfer}
\end{figure*}

\begin{figure*}[!t]
\centering
\subfloat[]{%
  \includegraphics[width=0.28\textwidth]{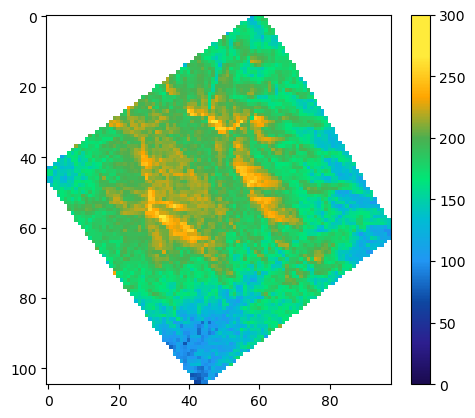}%
}
\subfloat[]{%
  \includegraphics[width=0.28\textwidth]{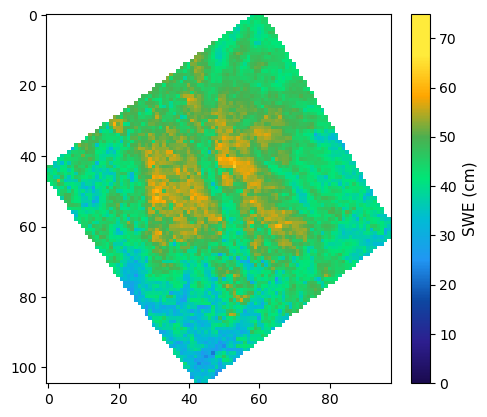}%
}
\subfloat[]{%
  \includegraphics[width=0.28\textwidth]{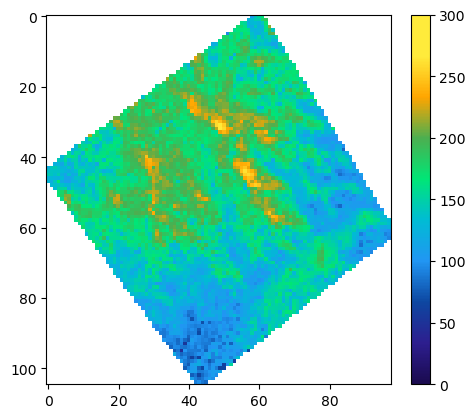}%
}
\caption{Spatial transfer results for BS2020 $\rightarrow$ MC2020. Panels show (a) lidar snow depth reference, (b) Retrieved SWE using physics-based approach \cite{oveisgharan2024snow}, and (c) MLP-predicted snow depth.}

\label{fig:mc2020_transfer}
\end{figure*}

\begin{figure*}[!t]
\centering
\subfloat[]{%
  \includegraphics[width=0.24\textwidth]{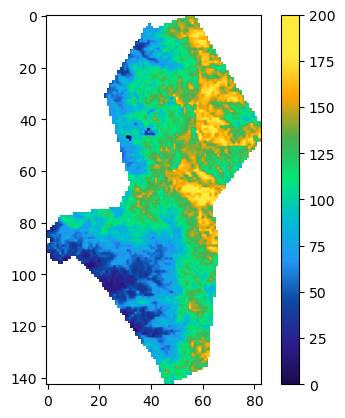}%
}
\hspace{0.5cm}
\subfloat[]{%
  \includegraphics[width=0.24\textwidth]{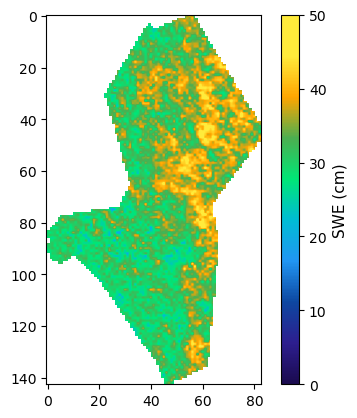}
}\hspace{0.5cm}
\subfloat[]{%
  \includegraphics[width=0.24\textwidth]{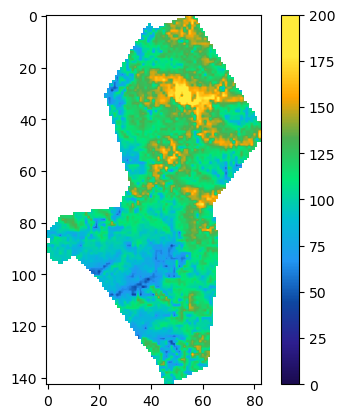}
}

\caption{Spatial transfer results for BS2020 $\rightarrow$ DC2020. Panels show (a) lidar snow depth reference, (b) Retrieved SWE using physics-based approach \cite{oveisgharan2024snow}, and (c) MLP-predicted snow depth.}
\label{fig:dc2020_transfer}
\end{figure*}

\begin{figure*}[!t]
\centering
\subfloat[BS2021]{%
  \includegraphics[width=0.21\textwidth]{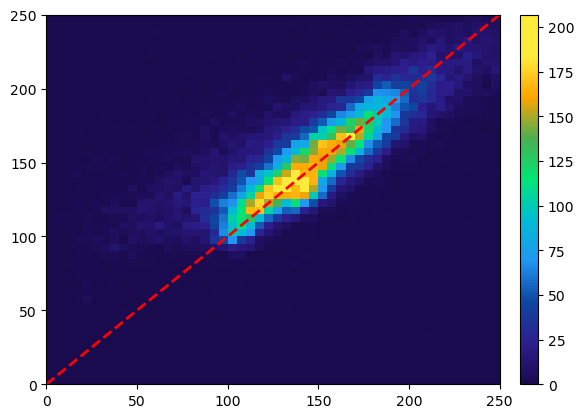}%
}
\subfloat[MC2020]{%
  \includegraphics[width=0.21\textwidth]{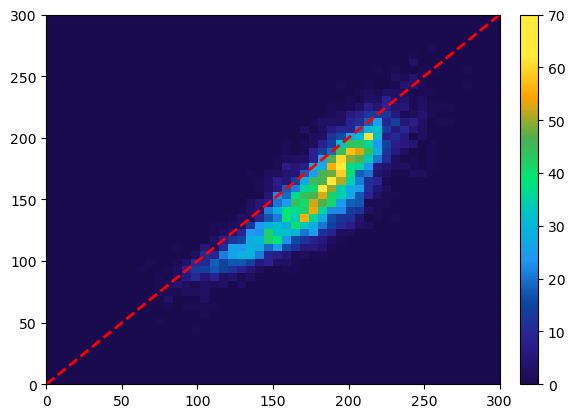}%
}
\subfloat[DC2020]{%
  \includegraphics[width=0.21\textwidth]{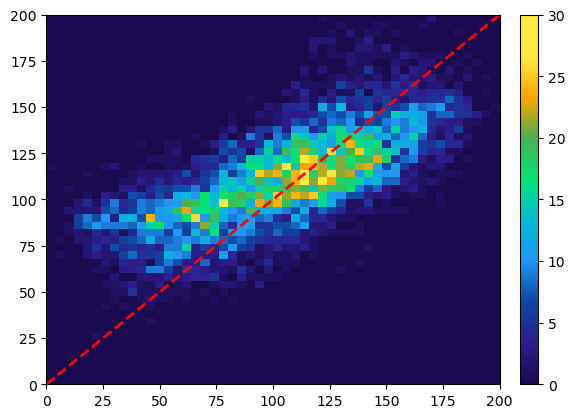}%
}

\caption{2D residual error histogram}
\label{fig:2D-error-hist}
\end{figure*}

\section{Methodology and data}

\subsection{Data}

We use the same Sentinel-1 interferometric datasets studied in~\cite{oveisgharan2024snow,oveisgharan2025using}. Sentinel-1 C-band repeat-pass SAR data acquired in Interferometric Wide (IW) mode were processed to obtain co-registered interferometric phase, coherence, and amplitude products at approximately 80\,m spatial resolution, such that each valid pixel corresponds to an $\sim$80\,m $\times$ 80\,m ground area. ASF’s On-Demand Processing system, Hybrid Pluggable Processing Pipeline (HyP3), is used to generate $6$-day interferometric products. 
Reference snow depth is obtained from NASA SnowEx airborne QSI lidar products, which provide high-resolution snow depth and vegetation height maps.

Four co-registered datasets over Idaho are considered: Banner Summit 2020 (BS2020), Banner Summit 2021 (BS2021), Mores Creek 2020 (MC2020), and Dry Creek 2020 (DC2020). BS2020 is used for model training unless otherwise stated, while the remaining datasets are reserved for temporal and spatial transfer evaluation.

\paragraph{Features and Problem Formulation} \label{sec:features}

Snow depth retrieval is formulated as a supervised per-pixel regression problem. For each valid Sentinel-1 pixel (approximately 80\,m $\times$ 80\,m), we construct a feature vector from interferometric observables and ancillary terrain descriptors and predict the co-registered lidar snow depth.

InSAR predictors include 12-day repeat Sentinel-1 unwrapped interferometric phase, temporal coherence, and radar backscatter amplitude time series.
For each pixel, these quantities are provided to the model as temporal stacks spanning 12 acquisitions, preserving time-dependent variability. Interferometric coherence is summarized using its temporal mean and standard deviation to capture signal stability while limiting dimensionality. We additionally include a cumulative line-of-sight (LOS) proxy formed by summing the LOS product across acquisitions.

Ancillary predictors describe terrain geometry and vegetation structure. It has been shown that these parameters impacts the SWE retrieval accuracy using InSAR data \cite{oveisgharan2025using}. These include local incidence angle and terrain attributes derived from a digital elevation model (DEM), elevation, incidence angle, slope, and aspect, together with lidar-derived vegetation height to account for canopy-related decorrelation effects.

The resulting per-pixel feature vector comprises $21$ channels: average unwrapped-phase, 12-day amplitude measurements, amplitude mean, two coherence summary statistics(mean and standard deviation), incidence angle, and four terrain/vegetation descriptors (slope, aspect, elevation and  vegetation height). This compact representation captures temporal interferometric behavior and static scene characteristics relevant to snow-depth retrieval.

Let $\mathbf{x}_p \in \mathbb{R}^{C}$ denote the resulting feature vector for pixel $p$ and $y_p$ the corresponding lidar snow depth. The learning objective is to estimate a mapping $f_\theta(\mathbf{x}_p)\approx y_p$ using supervised training data and to evaluate generalization under progressively more challenging conditions. Specifically, we consider four evaluation regimes: (1) an in-distribution setting where training and testing are performed on BS2020 using a held-out split; (2) temporal transfer from BS2020 to BS2021; (3) spatial transfer to a geographically distinct site with similar snow and terrain statistics (BS2020 $\rightarrow$ MC2020); and (4) spatial transfer under stronger distribution shift (BS2020 $\rightarrow$ DC2020). This progression isolates model behavior under matched conditions, temporal variability, and increasing spatial heterogeneity.

\subsection{Learning model}\label{sec:model}
Snow depth is estimated using a compact multilayer perceptron (MLP) operating on per-pixel feature vectors. The network comprises three fully connected hidden layers with widths $[128, 64, 32]$ and rectified linear unit (ReLU) activations, followed by a linear output layer that predicts snow depth. The architecture is intentionally low-capacity to encourage feature-driven generalization and to reduce overfitting to site-specific characteristics.

Training minimizes mean squared error with an 
$\ell_2$ regularization penalty (alpha) of $0.01$, using the Adam optimizer with an initial learning rate of $10^{-3}$ with adaptive scheduling. An internal 10\% validation split is used to monitor convergence, with early stopping applied using a patience of 15 epochs and a maximum training limit of 500 epochs. In practice, training converges well before the epoch limit. All experiments use a fixed random seed to ensure reproducibility.

The input feature vector combines acquisition-wise interferometric observables with terrain and geometry descriptors, forming a compact representation of radar signal behavior relevant to snow-depth retrieval. Each pixel-feature vector is paired with the corresponding lidar snow depth target, and the model learns a direct nonlinear mapping from interferometric measurements and ancillary variables to snow depth. This formulation enables controlled evaluation of temporal and spatial generalization without imposing explicit physical inversion constraints.

\section{Results}\label{sec:results}

Model performance is evaluated under the four regimes described in Section~\ref{sec:features}. For all experiments, predictions are compared against co-registered lidar snow depth using Pearson correlation and root-mean-square error (RMSE). 

\subsection{In-Distribution Baseline}

We first establish a matched-distribution reference by training and testing on BS2020 using an 80/20 pixel split. As shown in Table \ref{tab:baseline_results}, the model achieves strong agreement with Lidar-derived snow depth, with Pearson correlation of $0.91$ and RMSE of 0.15 (m) on test data,  indicating that the interferometric and terrain-based predictors contain sufficient information to recover snow depth when the training and testing distributions are aligned. This baseline serves as an upper-bound reference for the subsequent transfer experiments.

\subsection{Temporal Transfer}

Temporal generalization is evaluated by training the model on BS2020 and testing it on BS2021. Despite interannual variability, as shown in Table~\ref{tab:transfer_results}, the model maintains strong agreement with lidar measurements achieving Pearson correlation of $0.81$ and RMSE of $0.25$ (m), compared to  a Pearson correlation of $0.47 $ reported in \cite{oveisgharan2025using}. This result demonstrates that the learned mapping is stable across years when terrain and acquisition geometry remain consistent. 

This behavior is further supported by the 2D error histogram in Fig.~\ref{fig:2D-error-hist}(a), where predictions are tightly concentrated along the $x=y$ line, indicating strong agreement across the full snow-depth range. 
Visual results in Fig.~\ref{fig:bs2021_transfer} further confirm that the model preserves the dominant spatial structure of snow depth. As shown in the figure, the compact learning-based model successfully captures both the shallower snow depths on the eastern side of the scene and the pronounced north–south gradient of deeper snow in the central region compare to retrieved SWE using InSAR. In addition, the model reduces the discrepancy observed in the northwest corner, where the physics-based SWE retrieval indicates deep snow (yellow), while lidar measurements show only moderate snow depth (green). Note that both the compact learning-based model and the SWE retrieved from InSAR overestimate snow depth in very shallow snow regions (dark blue areas).

\subsection{Spatial Transfer: Moderate Shift}

To demonstrate the robustness of the trained model and its spatial generalization, we evaluate the same model trained on BS2020 on MC2020 dataset. Despite the geographic separation, as shown in Table~\ref{tab:transfer_results}, the performance remains strong achieving a Pearson correlation of $0.85$ and RMSE of $0.29$ (m), indicating that the model captures transferable relationships in the interferometric observables rather than memorizing site-specific patterns.

The 2D error histogram in Fig.~\ref{fig:2D-error-hist}(b) shows a systematic deviation below the $x=y$ line, indicating underestimation of snow depth and a bias in the mean prediction. This suggests that the model captures spatial variability but struggles to recover the absolute depth level at this site.

Visual results in Fig.~\ref{fig:mc2020_transfer} further confirm that the model preserves the dominant spatial structure of snow depth. As shown in the figure, the compact learning-based model successfully captures the shallower snow depths on the eastern side of the scene compared to the retrieved SWE using InSAR. However, as indicated by the 2D histogram, the model exhibits a systematic underestimation of snow depth.

This interpretation is further supported by the results in Table~\ref{tab:bias_effect}, where training on mean-centered targets significantly reduces RMSE while preserving correlation. In this setting, the mean is subtracted independently from both the training and test datasets. These results indicate that the dominant source of error is a global offset rather than a structural mismatch in the predicted spatial patterns. In practice, such offsets can be corrected using simple, low-complexity estimation procedures.
\subsection{Spatial Transfer: Strong Shift}

Finally, we evaluate the model trained on BS2020 on DC2020, which exhibits a considerably wider snow-depth range. As shown in Table~\ref{tab:transfer_results}, performance degrades relative to previous regimes. The 2D error histogram in Fig.~\ref{fig:2D-error-hist}(c) shows increased dispersion compared to the other cases, indicating reduced agreement between predictions and lidar measurements. Errors are particularly pronounced at low snow depths. This behavior is consistent with a stronger distribution mismatch between training and testing data, as well as limited representation of shallow snow conditions in the training set.

Visual results in Fig.~\ref{fig:dc2020_transfer} further confirm that the compact learning-based model successfully captures the very shallow snow depths on the western and southwestern parts of the scene (blue areas), whereas the SWE retrieved from InSAR fails to resolve these regions (green areas). However, as indicated by the 2D histogram, the model exhibits a systematic bias, with overestimation for very shallow snow depths.

\begin{figure}[!t]
\centering
\subfloat[]{%
  \includegraphics[width=0.22\textwidth]{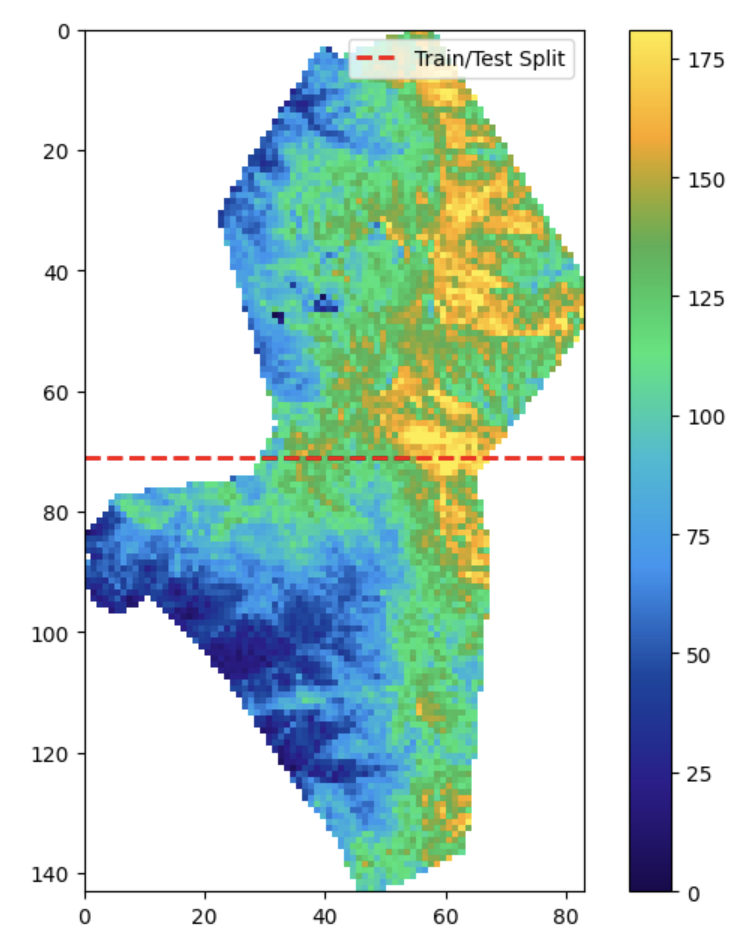}%
}
\subfloat[]{%
  \includegraphics[width=0.22\textwidth]{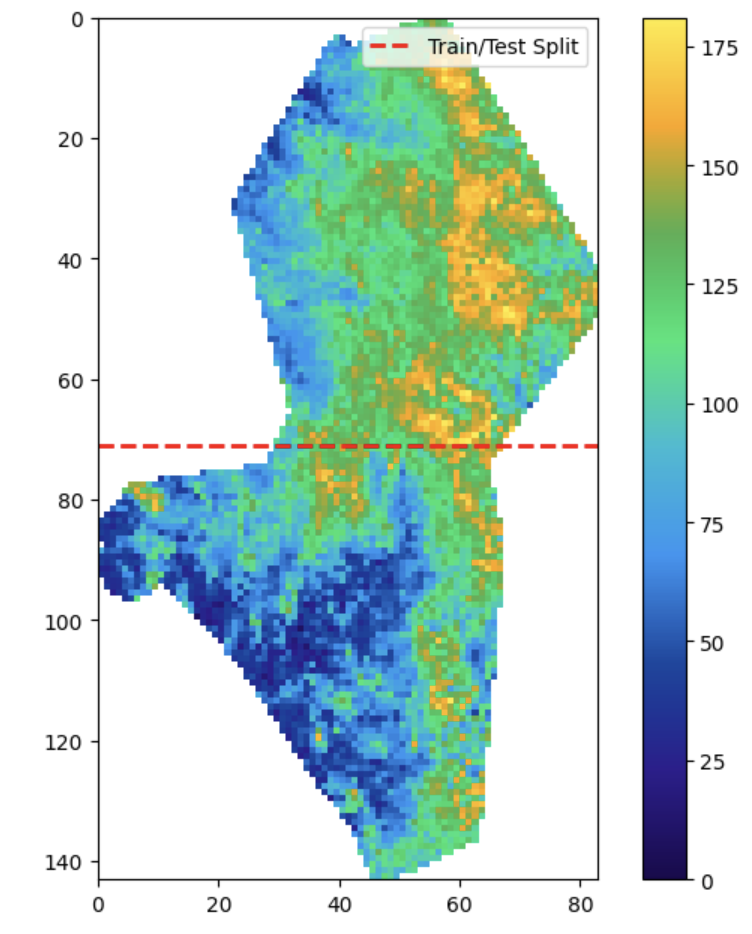}%
}

\caption{Spatial distribution of snow depth estimates for the DC2020 dataset. (a) LiDAR-derived snow depth and (b) model-predicted snow depth. The red dashed line denotes the train/test spatial split; the model was trained on the upper half of the domain and evaluated on the lower half. Predictions in the upper region of (b) reflect in-sample (training) performance, while predictions in the lower region reflect out-of-sample (test) generalization.}
\label{fig:dc2020_transfer_2}
\end{figure}

To further clarify the role of training-data mismatch, Fig.~\ref{fig:dc2020_transfer_2} presents results from a spatial transfer experiment in which the DC2020 dataset is partitioned into two halves. Upper half of the image is used for training and the lower half used for testing. The model achieves a Pearson correlation coefficient of $0.90$ with an RMSE of $0.14$ (m) on the training subset, and a Pearson correlation of $0.80$ with an RMSE of $0.23$ (m) on the held-out testing region. The panels show (a) the lidar-derived snow depth reference and (b) the MLP-predicted snow depth. These results indicate that when the training subset spans a sufficiently broad range of snow conditions, the model captures the spatial variability of snow depth and generalizes effectively to previously unseen regions.

\subsection{Discussion}

This work demonstrates that a compact learning-based model can directly map InSAR observables to snow depth and generalize across time and space. Compared to the physics-based SWE retrieval approach in~\cite{oveisgharan2025using}, which relies on explicit modeling assumptions, the proposed method achieves substantially higher correlation with lidar snow depth across all evaluation sites (Table~\ref{tab:transfer_results}). In particular, correlation improves from 0.47 to 0.81 for BS2021, from 0.66 to 0.85 for MC2020, and from 0.56 to 0.71 for
DC2020. The model would benefit from additional training data in shallow snow regimes. With the increasing availability of continuous NISAR and Sentinel-1 time-series observations, together with expanding lidar datasets, there is strong potential to further enhance model performance.

It is important to note that the physics-based approach in \cite{oveisgharan2024snow} estimates SWE rather than snow depth. As a result, the comparison with lidar-derived snow depth is not strictly one-to-one. In contrast, the proposed machine learning model is trained directly to predict snow depth, which naturally leads to improved agreement with lidar measurements in terms of Pearson correlation.
That said, the consistently high correlation and low RMSE achieved by the proposed method demonstrate its capability to reliably estimate snow depth from InSAR observations.

These results indicate that learning-based approaches can more effectively capture the relationship between interferometric observables and snow state, while remaining robust to temporal variability and moderate spatial domain shifts.



\begin{table}[!t]
\centering
\caption{In-distribution baseline performance for BS2020 using an 80/20 split.}
\label{tab:baseline_results}
\small
\begin{tabular}{lccc}
\hline
\textbf{Dataset} & \textbf{Pearson $r$} & \textbf{RMSE (m)} & $\mathbf{R^2}$ \\
\hline
Training (BS2020) & 0.93 & 0.14 & 0.86 \\
Testing (BS2020)  & 0.91 & 0.15 & 0.83 \\
\hline
\end{tabular}
\end{table}

\begin{table}[!t]
\centering
\caption{Temporal and spatial transfer performance comparison with prior physics-based SWE retrievals for models trained on BS2020.}
\label{tab:transfer_results}
\small
\begin{tabular}{lccc}
\hline
\textbf{Site/Year} & \textbf{Correlation} & \textbf{Correlation } & \textbf{RMSE (m)} \\
& \textbf{(from \cite{oveisgharan2025using})} & \textbf{(our model)} &  \textbf{(our model)} \\
\hline
 BS2021 &0.47 &0.81 & 0.25 \\
 MC2020 & 0.66 &0.85 & 0.29  \\
 DC2020 & 0.56 &0.71 & 0.28 \\
\hline
\end{tabular}
\end{table}

\begin{table}[!t]
\centering
\caption{Model trained  trained on BS2020 and tested on MC2020, with and without debiasing.}
\label{tab:bias_effect}
\small
\begin{tabular}{lcc}
\hline
\textbf{Method} & \textbf{Pearson $r$} & \textbf{RMSE (m)} \\
\hline
 Baseline (absolute depth)   & 0.85 & 0.29 \\
  Mean-centered (bias-corrected)  &  0.85 & 0.19   \\
\hline
\end{tabular}
\end{table}

\section{Conclusions}

This letter presents a compact learning-based approach that directly maps InSAR observables to snow depth and demonstrates strong temporal and spatial generalization across independent sites. The results show that the proposed method achieves substantially improved agreement with lidar measurements compared to physics-based retrievals, demonstrating that the deep learning framework can effectively estimate snow depth from InSAR-derived observables.

Beyond its current performance, the approach provides a pathway toward scalable, spatially extensive snow depth mapping across diverse snow-covered regions, capabilities that are difficult to achieve with lidar alone due to its limited spatial coverage and sensitivity to weather conditions. Furthermore, as continuous NISAR and Sentinel-1 time-series observations become increasingly available, together with expanding lidar datasets for training and validation, the model can be further refined and generalized. This creates a clear opportunity to extend snow depth estimation to broader regions and conditions, enabling improved large-scale monitoring of snow resources.

Future work will explore integrating physics-based constraints with learning-based models and leveraging spatial structure to further improve retrieval accuracy.

\section{Acknowledgements} 
The research was carried out at the Jet Propulsion Laboratory, California Institute of Technology, under a contract with the National Aeronautics and Space Administration (grant no. 80NM0018D0004).


\bibliographystyle{IEEEtran}
\bibliography{myrefs_igarss}

\end{document}